\documentstyle[12pt]{article}

\def\alwaysmath#1{{\ifmmode{#1}\else{$#1$}\fi}}
\def\he#1{\hbox{\alwaysmath{{}^{#1}}{\rm He}}}

\def\li#1{\hbox{\alwaysmath{{}^{#1}}{\rm Li}}}

\def\hii{H\thinspace{$\scriptstyle{\rm II}$}}
\def\etal{{\it et al.}~}
\def\beginapjbib{\begingroup \section*{\large \bf References}
   \parskip=.5ex plus 1.0pt
   \def\bibitem{\par \noindent \hangindent\parindent
      \hangafter=1}}
\def\endapjbib{\par \endgroup}
\begin{document}
\begin{titlepage}
\pagestyle{empty}
\baselineskip=21pt
\rightline{UMN-TH-1402/95}
\rightline{astro-ph/9508086}
\rightline{August 1995}
\vskip .2in
\begin{center}
{\large{\bf The Local Abundance of $^3$He: A Confrontation
Between Theory and Observation}} \end{center}
\vskip .1in
\begin{center}
Sean T. Scully$^1$, Michel Cass\'e$^2$,  Keith A. Olive$^1$,
David N. Schramm$^{3,4}$, James Truran$^3$,
and Elisabeth Vangioni-Flam$^5$

$^1${\it School of Physics and Astronomy, University of Minnesota}
{\it Minneapolis, MN 55455, USA}

$^2${\it Service d'Astrophysique, DSM, DAPNIA, CEA,  France}

$^3${\it Department of Astronomy and Astrophysics, Enrico Fermi Institute,
The University of Chicago, Chicago, IL  60637-1433}

$^4${\it NASA/Fermilab Astrophysics Center,
Fermi National Accelerator Laboratory, Batavia, IL  60510-0500}

$^5${\it Institut d'Astrophysique de Paris, 98bis
Boulevard Arago, 75014 Paris, France}

\end{center}
\vskip .5in
\centerline{ {\bf Abstract} }
\baselineskip=18pt

Determinations of the \he3 concentrations in Galactic matter
serve to impose interesting and important constraints both
on cosmological models and on models of Galactic chemical
evolution. At present, observations of \he3 in the solar system
and in the interstellar medium today suggest that the \he3
abundance has not increased significantly over the history
of the Galaxy, while theoretical models of Galactic chemical
evolution (utilizing current nucleosynthesis
yields from stellar evolution
and supernova models) predict a rather substantial increase
in \he3.
We consider the possibility that the solar \he3 abundance may
have been affected by stellar processing in the solar neighborhood
prior to the formation of the solar system. Such a discrepancy between
solar abundances and average galactic abundances by as much as a factor
of two, may be evidenced by
several isotopic anomalies. Local destruction of \he3 by a similar
amount could serve to help to reconcile the expected increase in the
\he3 abundance predicted by models of galactic chemical evolution.
 We find however, that the production
of heavier elements, such as oxygen, places a strong constraint on
the degree of \he3 destruction.
We also explore the implications of both alternative models of Galactic
chemical evolution and the stellar yields for \he3 in low mass stars,
which can explain the history of the \he3
concentration in the Galaxy.

\noindent
\end{titlepage}
\baselineskip=18pt
\def\la{~\mbox{\raisebox{-.7ex}{$\stackrel{<}{\sim}$}}~}
\def\ga{~\mbox{\raisebox{-.7ex}{$\stackrel{>}{\sim}$}}~}
\def\beq{\begin{equation}}
\def\eeq{\end{equation}}

\section{Introduction}

There is an inherent difficulty associated with the utilization
of the observed abundances
of D and \he3 to predict their primordial values.  Namely,
the connection between the primordial abundances of D and \he3
and their solar or present-day values depends sensitively on models
of galactic chemical evolution. In principle, measurements
of D in quasar absorption systems could dramatically help us to bridge
this gap, by providing directly the primordial abundance of D and hence
the value of the baryon-to-photon ratio, $\eta$, from big bang
nucleosynthesis (Walker \etal 1991). However,
recent measurements of this
kind (Carswell \etal 1994; Songaila \etal 1994;
Tytler \& Fan 1995) must be
viewed as preliminary, as the determined D abundance
in the two absorption systems
observed are not concordant with each other.
Until such measurements yield a single
consistent value for primordial D, we must continue to be guided by
models of galactic chemical evolution.

It has been well established that models of galactic chemical evolution,
consistent with the constraints imposed by element abundance
determinations, are capable of destroying significant amounts
of deuterium (Truran \& Cameron 1971;
Gry \etal 1984; Delbourgo-Salvador \etal 1985;
Vangioni-Flam and Audouze 1988; Vangioni-Flam, Olive \& Prantzos
1994; Vangioni-Flam \& Cass\'{e}, 1995). However,
as was recently discussed in Olive \etal (1995),
the problem rests not with the destruction of D, but rather with
the production of \he3.
Though \he3 is partially destroyed in massive stars, \he3 production
in low mass stars generally leads to a net increase in the \he3
abundance over the evolutionary history of the galaxy.
Observations of large \he3 enhancements in planetary nebulae
(Rood, Bania \& Wilson, 1992; Rood, Bania, Wilson \& Balser,
1995) support the conclusion that \he3
is indeed produced in low mass stars.
An excess of \he3 is difficult to avoid if low mass stars are strong
producers of this isotope, as indicated by the calculations of Iben and
Truran (1978) and more recently Vassiliadis and Wood (1993) and
Weiss \etal (1995). However, recent models with
increased mixing have been calculated, which both bring
the carbon and oxygen isotopic ratios to
their observed level in red giants and lead to a
net destruction of \he3 (Charbonnel 1994; Wasserburg \etal 1995;
Hogan 1995). At present, these models are quite preliminary,
and it is premature to draw a firm conclusion. However, if their
results are confirmed, \he3 will  be significantly less problematic,
unless the primordial D abundance is as high as observed by Songaila \etal
(1994) and Carswell \etal (1994); in this case we would still require
\he3 destruction factors in excess of what current calculations bear out
(Olive \etal 1995).

 Abundance
determinations of \he3 at the time of the formation of the solar system
seem to indicate that the solar \he3 abundance
 is very close to that of the primordial abundance.
Here, we will examine in some detail the
 possibility that the \he3 abundance in the
solar system may be depleted with respect to galactic
averages at the time of its formation.
We find, however, that the abundances of the heavier
elements, most notably
oxygen and neon, impose a strong constraint on the
degree of depletion of \he3.
We are therefore left with the following possibilities: either the
initial deuterium abundance is
low, D/H $\sim 3 \times 10^{-5}$ (we will
discuss rough lower limits imposed by models of chemical
evolution); or more dramatic changes are required in models of
chemical evolution, which have the effect of maintaining a rather
flat evolution  of \he3 with time
(we will show an example of this type of model below); or the stellar yields
of \he3 in low mass stars are lower than previously thought.

Because deuterium is converted to \he3 in the
pre-main-sequence phase of
stellar evolution, models without a significantly
depressed initial D abundance are subject to
problems with \he3 production. If the initial
abundance of D is rather
low (D $ \la 3 \times 10^{-5}$), the present day
\he3 abundance found in standard
models of galactic chemical evolution are not
excessive and are consistent
with the observed range of $1 - 5 \times 10^{-5}$ (Balser \etal 1994).
However, even in these
cases, there appears to be a problem with the
abundance of \he3 as measured in
meteorites, giving the pre-solar value of \he3.
That is, on the basis of chemical evolution models,
we expect more \he3 than is observed in the
solar system (Olive \etal 1995;
Galli \etal 1995; Tosi \etal 1995). In this paper, we will thus
consider in turn,
the possibilities that \he3 in the solar system
has been depleted with respect to the
galactic average; the accuracy of the measurement
of solar \he3; and what we
can learn from galactic chemical evolution.

\section{Solar Depletion of \he3?}

In principle, it is possible that the abundance of \he3 at the
time of the formation of the solar system does not
reflect the galactic average at that time.
It is necessary to consider the degree to which
element abundances in the solar system
were affected by the explosions of
supernovae in the solar neighborhood,
immediately prior to the formation
of the solar system (Reeves 1978; Olive \& Schramm 1982).
 This may be evidenced by several ``anomalous"
isotopes (of Carbon, Oxygen and Neon) seen in cosmic rays.
However, late contributions to the abundances of  $^{16}$O and $^{20}$Ne,
 which are produced solely in Type II supernovae, may render the
solar system isotopic ratios
of these elements anomalous. That is,
the solar abundances may not represent the true average galactic
abundance. For the oxygen and neon isotopes, these differences
may be as large as a factor of two.  We now examine the possibility
that the solar \he3 abundance may also not be representative of the
true galactic abundance.

{}From the observed anomalies in $^{26}$Al and
$^{107}$Pd (Lee 1979; and references therein) and more recently $^{41}$Ca,
and the short time scales associated with their
half-lives ($\sim 10^6$ years), the
element abundances in our solar system were probably
affected by at least one
supernova within that time period prior to formation.
Even a single supernova
explosion in a star forming region can have dramatic
consequences on the element
abundances of that region. As was argued by Olive \&
Schramm (1982), a handful of the first
few supernovae in an early OB association, is capable of
producing nearly the entire
observed solar abundance of $^{16}$O and $^{20}$Ne.
Thus we would expect that the
solar isotopic ratios such as $^{17,18}$O/$^{16}$O and
$^{22}$Ne/$^{20}$Ne may be
diluted with respect to the galactic average.
It is therefore of interest to question
whether or not the abundance of \he3 (which would be
depleted in the ejecta of
the first few supernovae in an association) is comparably depleted.

To deplete \he3, we must require that a significant
amount of material
in the solar neighborhood underwent stellar processing prior to the formation
of the solar system.  Let us suppose that a
fraction $f$ of the total initial gas of
the association went into stars prior to
the solar epoch.  It is then reasonable
to assume that a fraction $\sim 0.1f$ of the gas
went into stars with masses
greater than 10 $M_\odot$. (For example, with a
Salpeter (1955)  initial mass function
(IMF) $\phi(m) \propto m^{-2.35}$
between, 0.1 and 100 $M_\odot$, the fraction is
12\%; for a Scalo (1986) mass function
it may range from 5--15\% depending on the star formation rate (SFR)).
If we denote by $X_*$ the mass fraction of
some heavy element (such as O)
ejected by massive stars,
the total mass fraction of the element after the
explosions of stars more massive
than 10 $M_\odot$ which thus determines (and must be less than)
 the solar abundance, is then given by
\beq
X_f = {0.1fX_* + (1-f)X_i \over (1-f) + 0.1 f} < X_\odot
\label{xf}
\eeq
where $X_i$ is the initial mass fraction of the
element and we have assumed that $0.9f$
of the initial gas mass is still locked in stars.
To maximize our estimate of
\he3 destruction (as this will maximize our estimate for $f$),
we can assume that $X_i = 0$.  Solving for $f$, we have,
\beq
f < {X_\odot \over 0.1X_* + 0.9 X_\odot}
\eeq
For oxygen, $X_* \sim 0.1$ and $X_\odot \sim 0.01$, so that $f \la 1/2$.
Thus, we can cycle no more than 1/2 the mass of the
association through stars
prior to the formation of the solar system.

Although a significant amount of gas may be cycled
through stars, only a small fraction
of \he3 depleted gas can be released back into
the association.  If we take $X$
to represent \he3 in Eq. (\ref{xf}), and now
take $X_* = 0$ (which assumes that
\he3 is totally destroyed in massive stars),
and $X_i $ to be the primordial
\he3 mass fraction, we find
\beq
X_f = {(1-f)X_i \over 1 - 0.9f} \la .9 X_i
\eeq
This indicates that only about 10\% of the initial \he3 can be
destroyed, even though changes in the
heavy element abundances occur at a level of a factor of 2.

It is possible, of course, to further deplete
\he3 in the gas which forms the solar system
at the expense of excessive metal production.
Such overproduction of metals can perhaps be reconciled with
\he3 depletion, if the heavy elements could somehow be
expelled from the solar neighborhood.
As Lattimer, Schramm \& Grossman (1977) pointed out, the bulk
of the heavy element ejecta from supernovae can rapidly form into
dust grains.  These dust grains can behave like explosive ``shrapnel"
and penetrate regions exterior to the association.  This might allow
the association itself  and hence the solar
system, to fail to show a large heavy
element excess, even though the total heavy element enrichment would be part
of the integrated galactic enrichment. This assumes, of course,
that the entire association region is not totally disrupted by the
supernovae explosion prior to the formation of the solar system.
However, as was shown in Olive \& Schramm (1982),
a significant amount of oxygen
and neon is produced which should not be trapped in
grains. Because of the behaviors of these
elements, we believe that is unlikely that more than
about 10\% of the \he3 present
in the association could be destroyed before the
formation of the solar system.

An obvious recourse to resolving the problem of
the overproduction of \he3 at the solar epoch,
is to question the measurement of the solar
\he3 abundance. In the next section, we
examine the observational data on the solar abundances of D and \he3.

\section{D and $^3$He in Pre-solar Nebulae}

Because the crucial data in attempts to estimate the primordial D/H and
$^3$He/H
values come from solar system measurements, it is useful to
examine critically
the origin of these abundances and to attempt to provide an accurate
estimate of their
uncertainties.

The determination of the solar abundances of D and \he3 involves
 $^3$He/$^4$He measurements both in
meteorites and also directly
in the solar wind.  Direct D/H measurements are
irrelevant
for the Sun, since D is completely burned to $^3$He in the solar
convective
zone. Moreover, D/H measurements are
difficult to interpret in planetary bodies (Earth, Jupiter, etc.);
because D preferentially enters molecules relative to H,
abundance determinations thus require a knowledge of
complex chemical fractionation histories.
However, because essentially all
primordial D has been burned to $^3$He in the solar convective zone and
because
the convective zone is not hot enough to burn $^3$He, the solar wind
measurement of $^3$He/$^4$He provides a measurement of the pre-solar
abundance of ${{\rm D + ^3He}
\over ^4 He}|_\odot$ by number.
Solar wind measurements, made using foil
collectors during Apollo lunar
missions, yielded values for $^3$He/$^4$He ranging from 4 to 5.5  $\times
10^{-4}$. Geiss and Reeves (1972) and Bochsler \& Geiss (1989)
(see also Geiss (1993) for a recent review)
 argue that the variation can be corrected for, and that the best
solar
wind ratio is ${\rm ^3 He/ ^4 He \vert _{sw} = 4.1
\pm 1 \times 10^{-4}}$
(where the error is statistical). This is in good agreement with the low
temperature component emitted by carbonaceous chondrites in step-wise
heating
experiments (Black 1972; Weiler \etal 1991),
for which ${\rm ^3 He/ ^4 He \vert _{sw} \simeq 4.5
\pm 1 \times 10^{-4}}$, and also with the ISEE-3 solar wind data
(Coplan \etal, 1984), which yields $4.4 \times 10^{-4}$.
However, some fractionation
in all
the solar wind $^3$He/$^4$He measurements cannot
be excluded, which would add
an
additional systematic error to the above value.
The extreme value for $^3$He/$^4$He
observed for the Apollo solar wind measurement of
 $5.5 \times 10^{-4}$ can not be excluded as a central value, hence
a systematic uncertainty
of $1.4 \times 10^{-4}$ for $^3$He/$^4$He cannot be
excluded at present. The most recent measurement of
$^3$He/$^4$He  in the solar wind
from the over-the-solar-pole measurements
made with the
SWICS
instrument on the ULYSSES spacecraft gives \he3/\he4 $= (4.4 \pm 0.4) \times
10^{-4}$ (Bodmer \etal 1995).

The pre-solar $^3$He/$^4{\rm He}\vert_\odot$ ratio is thought to be best
measured
in meteorites.  Initially, Black (1971) proposed
that the high temperature
component emitted by step-wise heating experiments using carbonaceous
chondrites
(see also Eberhardt 1974) was the primordial component. $^3$He/$^4 {\rm
He} \sim 1.5 \times 10^{-4}$. However, Weiler \etal (1991)
 have argued that this
high temperature component is dominated by gas
trapped in pre-solar grains
(diamonds) which formed in locations far removed from the solar system.
Weiler
\etal propose that another gas component known as ``Q" is a better
candidate
for the primordial component.  Fortunately,
the difference in $^3$He/$^4$He
between the high T carbonaceous chondrite component and Q is relatively
small
\beq
^3 He/^4 He \vert_Q = 1.6 \pm 0.04 \times 10^{-4}
\eeq
However, a potential interpretational (systematic) error persists,
since
neither Q nor diamonds nor the high T carbonaceous
chondrite component has
been
unequivocally proven to represent $^3$He/$^4{\rm He}\vert_\odot$. Taking
$^3$He/$^4{\rm He}\vert_q$ as
$^3$He/$^4{\rm He}\vert_\odot$, but allowing for
systematics to include the range of relevant meteoritic $^3$He/$^4$He
values,
yields
$^3$He/$^4{\rm He}\vert_\odot = 1.6 \pm 0.04
\pm 0.3 \times 10^{-4}$. The
pre-solar
D is estimated by subtracting
$^3$He/$^4{\rm He}\vert_\odot$ from the SWICS solar
wind
value.  To convert to ratios relative to
 hydrogen requires multiplying by
the
number ratio of ${\rm ^4He/H \vert_\odot}$, which is
 estimated to be $0.09 \pm
0.01$
(note, this is 10\% lower than that used by Geiss 1993)
from the best fit
solar
model $Y = 0.27$ (Turck-Chieze \etal 1988;
Bahcall and Pinnsoneault 1992) with
metallicity Z = 0.02. This yields
\beq
({{D + ^3 He} \over H})_\odot = 4.1 \pm 0.6 \pm 1.4 \times 10^{-5}
\eeq
and
\beq
{{^3 He} \over H} \vert_{\odot} = 1.5 \pm 0.2 \pm 0.3 \times 10^{-5}
\eeq
and thus
\beq
{{D} \over H}_{\odot} = 2.6 \pm 0.6 \pm 1.4 \times 10^{-5}
\eeq

This latter number is in reasonable agreement with the ${\rm {HD} \over
{H_2}}
= 1 - 3 \times 10^{-5}$ ratio measured in Jupiter (Smith, Scherpp \&
Barnes,
1989). Although planetary D ratios are subject to chemical
fractionation, this is minimized for HD on Jupiter, since the bulk of the
deuterium and hydrogen is in HD and H$_2$
 there. However, molecular line blanketing does still
allow
for significant systematic errors.
For this reason, Jupiter is still not the best
source
for a solar system D determination, but it does provide a consistency check.

\section{Chemical Evolution}

The solar system abundance of $^3$He is thus seen to be approximately
a factor of two lower than predicted by
even the more optimistic models of
Galactic chemical evolution, which tend to yield abundance ratios
at least as high as $3 \times 10^{-5}$
for \he3/H, when \he3 production in lower mass
stars is included (Olive \etal 1995).
In what follows, we will look at three different
approaches to resolving the
problem of excess solar \he3. We first consider
possibilities for which the
primordial value of D/H is low.  A low initial
D/H lowers \he3/H, as there is less
D to be converted to \he3 in the pre-main-sequence
evolution of stars. However, as we
will show, one can not take arbitrarily low values
of D/H (of course D/H
is always bounded from below by the ISM measurements
of D/H yielding
D/H = $1.6 \pm 0.09 {}^{+0.05}_{-0.1}$
(Linsky \etal 1993, 1995)), since some
amount of deuterium destruction necessarily
accompanies the production of heavy elements
in the galaxy. We then consider ``higher"
values of D/H, which require
some dramatic changes to simple models of
chemical evolution, such as an
increased production of massive stars in the
early galaxy as well as metal enriched
outflow. We will also examine some remaining
alternatives regarding the stellar
production of \he3.  Note however, that there may be a
quite disturbing dispersion of D/H in the local ISM which would
complicate the analysis (Vidal-Madjar 1991; Ferlet 1992, Linsky
private communication)

As was noted earlier, the questions concerning
high verses low D/H may become moot, if the determinations of primordial D/H
in quasar absorption systems yield a single
consistent value. To date there are three
measurements of D/H in quasar absorption systems.
Two (in the same system) yield
a high value for D/H $\approx 1.9-2.5 \times
10^{-4}$ (Carswell \etal 1994; Songaila
\etal 1994), while the third (in a different system
 yields a significantly lower value, D/H $\approx 1-2 \times
10^{-5}$. It is clear that, on the basis
of these measurements, we
can not with confidence claim any knowledge
of the primordial abundance of
deuterium.  Indeed, it has been argued ( Levshakov \& Takahara 1995)
that measurements of this type may
not be able to determine D/H to better
than an order of magnitude. In other
words, they would expect a large dispersion in the observational data.
Is this what we are seeing?

Interestingly enough, the two values for D/H identified above are
in some respects both
beneficial and detrimental to big bang
nucleosynthesis. The high value of D/H
corresponds to a value for the baryon-to-photon
 ratio $\eta \simeq 1.5 \times
10^{-10}$ (Walker \etal, 1991). Consequences
of this high D/H were recently discussed in
Vangioni-Flam \& Cass\'e (1995).  With regard
to the other light elements produced
in big bang nucleosynthesis, the low value
for $\eta$ corresponds to a \he4 mass
fraction $Y_P \simeq 0.23$, which is in
remarkable agreement with what one expects
from the data on \he4 from extragalactic
\hii~regions (Olive \& Steigman 1995; Olive \& Scully 1995).
\li7/H is predicted to be around $2 \times
10^{-10}$ which is also compatible
within errors, with recent data (Molaro \etal 1995).
The problem occurs with the evolution
of \he3,
when \he3 production is included
(note that models of chemical evolution
can be constructed which can account
for the necessary D/H destruction in this case).
In Olive \etal (1995), it was found that
the abundance of \he3 at the time of solar system formation
could be high by as much
as a factor 10. Even in the absence of
\he3 production, it was found that massive
stars were required to destroy at least 90\%
of their initial D + \he3, in order to
reproduce the solar and ISM values of \he3.
This amount of destruction is excessive,
even for the most massive stars (Dearborn, Schramm \& Steigman, 1986).

On the other hand, the low value of D/H
between 1 and 2 $\times 10^{-5}$ corresponds
to a value of $\eta \approx 7 - 9 \times 10^{-10}$.
In contrast to the high
D/H case, we would expect
a much milder problem with \he3 (to be discussed below).
However, now the \he4 mass
fraction is predicted to be $Y_P > 0.249$, a value larger
than most of the \he4 measurements
(Pagel \etal, 1992; Skillman \etal 1995)
in extragalactic \hii~ regions, which already
contain some non-primordial \he4. (However, again, possible systematic
errors can not be excluded Copi \etal, 1995a; Sasselov \&
Goldwirth, 1995.)
In addition, \li7/H is expected to be $> 5
 \times 10^{-10}$ requiring a significant
amount of \li7 depletion, contrary to what
one expects (Steigman \etal 1993) from the
positive measurements of \li6 in halo stars
(Smith Lambert \& Nissen, 1992;
Hobbs \& Thorburn, 1994).
Furthermore, as we will next show, a minimal
 amount of D destruction is
demanded for consistency with the observed level of
heavy element production in the Galaxy.
A completely flat evolution for D is probably excluded on these grounds.

The classical constraints on galactic evolution are characterized
by varying degrees of stringency.
Among these, the trends in [Fe/H] with time are easily satisfied,
since the age-metallicity
relation suffers from a large dispersion over the observed age
range (Edvardsson \etal 1993; Nissen 1995). The [O/Fe] vs [Fe/H]
relationship is mainly sensitive to the stellar yields and not to the
different histories of star formation (assuming a constant IMF).
The metallicity distribution of
disk stars is far from being definitely established. Indeed, much
work is needed before a clear picture of the metallicity distribution
can be reached (e.g. Olsen 1994; Cayrel, private communication).
Information on metallicities,
ages and kinematics, with the same high
accuracy as obtained by Edvarsson \etal (1993), is needed for a much
larger stellar sample. Moreover, Grenon (1989, 1990) remarks that
the radial migration of stars in the Galaxy can blur the local
metallicity distribution.

Other global characteristics which should be considered are the gas
fraction, $\sigma$, the overall metallicity, Z, and individual abundance ratios
(Fe/H, O/H,...) at solar birth and in the present ISM.
 To the list of constraints, we must also add the D/H and \he3/H ratios at
solar birth and at present time, in relation to the  primordial value.
Indeed, since primordial nucleosynthesis is much more constrained
than galactic evolution, it is reasonable to harmonize the second  to
the  first, and not the contrary (as has sometimes been done recently).

Many models have been proposed to follow the chemical evolution
of the Milky Way, invoking, for example, a prompt initial
enrichment ( Truran \& Cameron 1971), infall
of primordial material (Timmes \etal 1995; Fields 1995), metal
enriched infall originating from the halo (Ostriker and Thuan 1975),
and early massive star formation ((Larson 1986; Wyse and Silk
1987).
Studies of galactic chemical evolution remain in their infancy, however,
since we do not yet have good theories of galaxy formation
and star formation. It would be
unwise, for the sake of simplicity, to limit the investigation to
``classical" models under the pretext that they have been widely
used. In effect, if the high primordial D/H ratio is confirmed, special
models leading to a strong D destruction avoiding overproduction of
\he3 and Z will be required.

An alternative way of looking at variations from the galactic mean has
been carried out by Copi, Schramm \& Turner (1995b) looking at
the stochastic variations from galactic evolution models.
Their conclusions concerning the allowed range of primordial
D and \he3 are similar to, and compatible with, those we discuss here.

\subsection{Low D/H}

We will first explore the possible consequences of a very low primordial
value of D/H and examine the extent to which a low D/H could explain the
apparent flatness of the \he3/H evolution in the Galaxy.
We begin by estimating the minimum possible amount of D/H destruction.
In simplified models of galactic chemical evolution, it is possible to
derive some analytic relations between abundances, yields, the
gas fraction, and the IMF, if one assumes the instantaneous
recycling approximation (that is, that the enriched mass that is ultimately
to be ejected from a star is incorporated into the ISM at the time of
formation of the star, in contrast to its appropriate delayed entry
at the end of the star's lifetime).
 Indeed, the degree to which
deuterium is destroyed can be expressed simply by (Ostriker and Tinsley 1975)
\beq
{\rm \frac{D}{~D_p}} = \sigma^{R/(1-R)}
\label{D}
\eeq
where $\sigma$ is the gas mass fraction and the return fraction, $R$,
is given by
\beq
R = \int_{M_1}^{M_{sup}} (M-M_{rem}) \phi(M) dM
\label{R}
\eeq
In (\ref{R}), $M_1$ is the main-sequence turnoff mass (normally a function of
time), $M_{sup}$ is the upper mass limit for star formation, and
$M_{rem}$ is the remnant mass.

It is also possible to express the metallicity in terms of the gas mass
fraction
and the yields of metals in stars (Searle \& Sargent 1972)
\beq
Z = {P_Z \over (1 - R)} \ln \sigma^{-1}
\label{Z}
\eeq
where
\beq
P_Z = \int_{M_1}^{M_{sup}} ({M_Z \over M}) M  \phi(M) dM
\eeq
and $M_Z/M$ is the mass fraction ejected in metals. Equations (\ref{D})
and (\ref{Z})
can be combined yielding
\beq
{\rm {D \over ~D_p}} = e^{-{ZR \over P_Z}}
\eeq
As one can see from Eq. (\ref{D}), a low primordial value for
D/H, will require a small return fraction, $R$. In principle, one can easily
adjust the IMF to yield a small value for $R$.  However, because
of the similarity in the definitions of $R$ and $P_Z$, their ratio
is almost independent of the details of the IMF.
  Thus ${\rm D/D_p}$ near unity, implies a
metallicity much less than solar.

The interdependence between deuterium and
metallicity can be seen in Figure 1.
In order to reach  solar metallicity at the
time the solar system formed, we require a deuterium
destruction factor of at least 1.6, implying that
D/H$_p \ga 2.5 \times 10^{-5}$.
We note that this factor is somewhat dependent upon
the assumed yields for the heavier elements.   For example,
this limit was obtained using the stellar yields of Woosley \& Weaver (1993),
whereas
had we used the yields of Maeder (1992), which allow for more
heavy element production in the mass range from 9 - 11 M$_\odot$,
the minimum destruction factor could be lowered to about 1.3.
It is worth noting that, beyond the uncertainties of the yields which
are essentially related to those associated with
the $^{12}$C($\alpha,\gamma$)$^{16}$O
reaction rate, the lower mass limit of the stellar progenitor of
the Type II supernovae which synthesize the heavy elements
is influential because of the preference in the IMF towards lower mass
stars. Indeed the limit is greatly increased as the slope of the IMF
is decreased.  All of these effects can be seen in Figure 1, where
we have plotted
(for various choices of the parameters which govern the
SFR) the  metallicity at the solar
epoch in units of solar metallicity, Z/Z$_\odot$, as  a function of
the ratio of the {\em present} deuterium abundance to the primordial one,
thus indicating the total deuterium destruction factor for a variety
of  galactic
evolution models. We have chosen
 a SFR proportional to the mass in gas,
and an IMF, $\phi(m) \propto m^{-2.7}$; shown here by the upper
set of points denoted by circles (yields from Maeder (1992)) and crosses
(yields from Woosley \& Weaver (1993)). For the lower set of points,
a steeper IMF, $\phi(m) \propto m^{-3}$ was chosen. In each case, the lower
mass limit of the IMF was 0.4 M$_\odot$ (lowering this choice to 0.1 M$_\odot$
would further lower the curves).
It is important to note that, when \he3 production is taken into account, even
the modest deuterium destruction factor (of 1.6),
 yields an overproduction
by about a factor of 2 in solar \he3.

\subsection{Higher D/H}

In this section, we will consider an alternative to low primordial
D/H and rely on more distinctive models of galactic chemical
evolution to resolve the problem concerning the  solar \he3 abundance.
As we have stated earlier, the choice of a higher value for primordial D/H
alleviates some of the pressure in matching the BBN calculations to the
observational determinations of \he4 and \li7.
Clearly, the higher the value we choose
for primordial D/H, the more difficult it
will be to keep \he3 under control.
We choose specifically the value D/H$|_{\rm p} = 7.5 \times 10^{-5}$
which corresponds roughly to the \li7 trough and is in modest agreement with
\he4 (at the 2 $\sigma$ plus systematics level).

The models we consider below specifically involve mass outflow.
Open galactic models have been considered in the past
(Tinsley 1980, Tosi 1988), but infall has been invoked more often
than outflow. Formally, the two reverse processes are included in
the general formalism of chemical galactic evolution (Tinsley 1980)
and cosmochronology (Cowan \etal 1989).
There is clear evidence for galactic winds in external
galaxies, even for spirals (Wang \etal 1995), and particularly those
experiencing bursts of star formation, whereas evidence for significant infall
of extragalactic matter are meager (Murphy \etal 1995).
Of course this reflects only constraints arising from
the present state of the solar vicinity, and
proves nothing about the early galaxy.

Outflow has its own merit: as we will see below, it will help to explain
very high destruction factors of D
(Vangioni-Flam \& Cass\'{e} 1995) while, if necessary, avoiding, an
overproduction of metals. At the same time it could reduce the rise of the
\he3/H ratio.
De Young \& Heckman (1994) proposed that energy from supernova
explosions and stellar driven winds results in blowing portions of the
ISM containing enriched material out of the galaxy.
Different kinds of outflows can be imagined,
which vary considerably in their
durations, intensities and compositions. For our
present purposes, it is sufficient
to distinguish whether the outflowing matter consists solely
of the ejecta of massive stars, or rather whether it is composed
of normal ISM material
being blown out by supernova driven winds. In Cass\'{e} \etal (1995),
we will return to these distinctions in greater detail.

The three specific ingredients that must be added to canonical galactic
evolutionary models in order to obtain significant D destruction
without the overproduction of \he3 are: (i) an
early phase of massive star formation, which presents the advantage
of destroying  D and \he3 rapidly; (ii) a galactic wind related to the
corresponding SNII rate, which limits the rise of Z and \he3, leading
to an even more pronounced  decrease of D; and (iii) possible
modifications of stellar models, leading to an efficient destruction
of \he3, especially in low mass stars.

One way we have found
in which the solar value of \he3 may be lowered is to assume that the IMF
prior to the formation of the solar system was skewed more toward
massive star formation.  The presence of fewer lower mass stars
reduces \he3 production,
while more massive stars ultimately return only a fraction of the \he3
present during the
pre-main sequence phase to the ISM.
  We therefore
consider
models which begin with an IMF favoring more massive stars
early on galactic history but resemble a more normal IMF at later
times.

The problem that we immediately encounter is that the emphasis on
more massive stars results in an overproduction of heavy elements,
 such as $^{16}$O.  We have found that the (closed box) models
which are most successfully in keeping \he3 flat, while
 destroying enough D, also overproduce
$^{16}$O by a factor of $\sim$10. This problem
is alleviated by including outflow.
Indeed,
McCray \& Snow (1979) have shown that supernovae can generate ``chimneys,"
which can directly transport much of the
heavy element rich supernova debris out
of the galaxy.
We have therefore included ``enriched''
(relative to the ISM) outflow in our models, both to help solve the heavy
element overproduction problem and to obtain a
flatter \he3/H evolution.  In order to simulate this effect,
we have incorporated outflow into our models at a
rate proportional to the rate of ejection of materials from supernovae.

Since massive stars can lose large amounts of \he3 depleted outer material
 via winds before they explode, it is certainly possible
for them to deplete \he3 in their surrounding
ISM material and  eject their metals out of the Galaxy.
We allow the outer (hydrogen) envelope of the star which is
deficient in \he3 to return via winds to the ISM.
Then, in order to maximize the possible effect of an outflow which is tied to
the ejecta of exploding stars (M $>$ 8-10 M$_\odot$),
 a fraction
of the core is then expelled from the Galaxy.
Such models provide a natural way to understand the heavy element
abundances in the X-ray gas observed in clusters of galaxies
(typically, the heavy element enriched outflow is produced by elliptic
galaxies, (see e.g. Elbaz, Arnaud, Vangioni-Flam 1995)).
It might also be noted that early expulsion of metal rich supernova ejecta
is even easier in merger models,
where the early galactic building blocks have a lower mass.
Due to the epoch of more massive star formation at earlier
times, D is generally very efficiently destroyed in these models.

We should note at this point that our assumptions concerning winds from
massive stars prior to the supernova stage may be inappropriate at
early galactic epochs. The rate of mass loss is generally expected to
be dependent upon the initial metallicity of the star (Maeder, Lequeux,
\& Azzopardi 1980; Maeder \& Meynet 1994). Expectations from
theoretical studies are generally consistent with e.g. trends in the
frequency of Wolf-Rayet stars, as inferred from studies of the
Magellanic Clouds (Massey \etal 1995). This suggests that the fraction
of the \he3 depleted outer envelopes of massive stars that is
returned to the ISM via winds (prior to supernova-triggered mass ejection)
in low metallicity populations may be significantly reduced.
We stress however, that our aim here is to see how efficiently the
evolution of the \he3 abundance can be held relatively flat over the
history of the Galaxy.  As we will see, we find only modest success
despite rather poignant assumptions.

An obvious observational constraint on our choice of an IMF that is
skewed toward more massive star production
at early times, $\phi(m) \propto m^{-(1.25 + O/O_\odot)}$,
is provided by its consistency with the present day IMF that
results from our model.  Fig. 2 shows a comparison of the observed and
modeled present day IMF.
The observed values are taken from Scalo (1986) and are
in good agreement with our model for the more massive stars.  This is as
expected,  since the more numerous massive stars formed early on have long
since
died out.

  It is be useful to compare the results we have obtained here with those of
our previous study, in which we considered a more
standard model. In model 1 of Olive \etal (1995),
we chose a SFR, $\psi = 0.25\sigma$, with an IMF,
$\phi(m) \propto m^{-2.7}$ between 0.4 and 100 M$_\odot$.
The \he3 abundance at the solar formation epoch (taken to be at $t = 9.4$ Gyr)
was \he3/H = 5.7 $\times 10^{-5}$, rising to 8.8 $\times 10^{-5}$
today. Infall was not included.
For a primordial ratio D/H = $7.5 \times 10^{-5}$,
the results for D/H and \he3/H as a function of time,
for a model with an IMF which
favors massive stars early,
and contains enriched outflow in which the rate is proportional to
the ejection rate, is compared with model 1 of Olive \etal (1995)
in Figure 3.
The outflowing gas contains only material below
the outer envelope, while the latter \he3 depleted material is returned to the
ISM.  In this model, the SFR is $\psi = 0.26 M_{\rm gas}$,
and the fraction of outflowing gas is 90\% of the supernova ejecta.
The IMF is now extended down to 0.1 M$_\odot$, to help keep
the metallicity and gas mass fraction reasonably low. We view this as a rather
extreme model, in which a considerable amount of enriched material
has been expelled from the galaxy.  Indeed, we impose a limit, arising from the
observed metallicity of hot X-ray gas in clusters,  on the amount of
metals expelled by outflow to be less than 20 times the amount of metals
in the galaxy. This imposes a constraint on the fraction of outflowing gas
(90\%
in this case).

As one can see from Fig. 3, our present model, which is based on
an IMF skewed towards massive stars early on and contains
enriched outflow, reduces the abundance of \he3 by a factor of about 2,
relative to the standard case with a normal IMF and no outflow.
Parameters of the model have been chosen such that the degree of deuterium
destruction is comparable (and agrees with the data) in the two cases.
However, although the present \he3 abundance is acceptable
\he3/H$|_o = 5.1 \times  10^{-5}$, the solar abundance is still
high by a factor of slightly over two; \he3/H$|_\odot = 3.7 \times 10^{-5}$.
While this represents a definite improvement, it can not be regarded as a
solution to the problem. Although it appears from Figure 3, better
agreement with the solar data is possible if one assumes a lower
time for the  formation of the solar system, the model must be adjusted
to destroy D on a faster time scale. For example, with $\psi = .34M_{\rm gas}$
the evolution of deuterium matches the solar
(and present-day) observations at $t =
6$ Gyr (corresponding to an age of the Galaxy of 10.6 Gyr), but now
\he3/H$|_\odot$
$\simeq 3.1 \times 10^{-5}$ and the present abundance is 4.9 $\times 10^{-5}$;
a further  improvement, but solar \he3 is still too high.

Of course as is well known, the problem concerning \he3 is also
alleviated somewhat by going to higher values of $\eta$.
In model 3 of Olive \etal (1995), we assumed a primordial abundance of
deuterium of D/H = 3.5 $ \times 10^{-5}$. In this case
\he3/H$|_\odot = 3.4 \times 10^{-5}$, an overproduction by a factor greater
than two. The present \he3 was also slightly high, \he3/H = $6 \times
10^{-5}$.
In models with outflow as described above, these numbers are reduced
to \he3/H$|_\odot = 2.3 \times 10^{-5}$ and \he3/H = $3.2 \times
10^{-5}$ today.

Before, we move on, we wish to stress that the problems concerning \he3
that we are discussing here, prevail only because we are including
the production of \he3 in low mass stars. When such production is ignored
there is no problem in matching the solar and ISM data for both
D and \he3 in models of these types as was shown by Vangioni-Flam,
Olive \& Prantzos (1994).  The crises in big bang nucleosynthesis claimed
by Hata \etal (1995) is only a crises because of the limit on the degree
of \he3 destruction they allowed.  Although the final \he3 abundance
in a given star relative to the initial D + \he3 abundance, usually
called $g_3$, is always larger than 0.25 as assumed by Hata \etal,
even simple models such as the type considered here (without outflow)
and in   Vangioni-Flam,
Olive \& Prantzos (1994) have an effective $g_3$ which is lower than
0.25 vitiating the purported crises.

\subsection{Alternatives}

A critical consideration with regard to the establishment of any
realistic constraints on cosmological D and D + $^3$He is that
associated with $^3$He production in low mass stars. Essentially
all early estimates of D and $^3$He constraints on cosmology
(see, e.g., Truran \& Cameron 1971; Rood, Steigman, \& Tinsley
1976) were based upon the stellar evolution models of Iben (1967 a,b),
for which analytical fits to the detailed model characteristics
were subsequently provided by Iben \& Truran (1978). The problem of
$^3$He then is simply the fact that, with the use of the Iben \&
Truran (1978) prescriptions, $^3$He production in stars in the mass
range $\sim$ 1-3 M$_\odot$ is sufficient to overproduce $^3$He in
Galactic chemical evolution models (Olive {\it \etal} 1995;
Galli \etal 1995; Timmes
\& Truran 1995), relative both to the solar system value of $^3$He and
to the $^3$He concentration in the ISM at the present time (Balser
{\it \etal} 1994). Further strong confirmation of this behavior has
been provided by recent stellar evolution calculations (Vassiliadis
\& Wood 1994; Weiss \etal 1995). It would seem to be necessary either to
utilize rather extreme assumptions regarding the history of our
Galaxy or to identify some significant problem in stellar evolution
theory.

An interesting recent paper by Wasserburg, Boothroyd, \& Sackmann
(1995) has called attention to the fact that the long-standing
problems associated with understanding both low $^{12}$C/$^{13}$C ratios
in low mass red giant branch stars and low $^{18}$O/$^{16}$O ratios
in asymptotic giant branch stars can be resolved, with the assumption
of the occurrence of deep circulation currents extending below the
bottom of the standard convective envelope. A concomitant of this
process of ``cool bottom burning'' is the destruction of $^3$He. In
particular, for the case of a 1 M$_\odot$ star, their models predict
that after a 1$^{st}$ dredge-up $^3$He enhancement of a factor $\sim$ 6,
cool bottom processing acts to reduce the $^3$He concentration by a
factor $\sim$ 10, yielding a net \underbar{depletion} of $^3$He by a
factor $\sim$ 2. If this model is indeed correct, this would aid substantially
in the problem of $^3$He overproduction, with which we are so
concerned in this paper.

To test the effect of such a reduction in the \he3 yields,
we incorporated the results of Wasserburg \etal (1995) by lowering the
Iben \& Truran (1978) yields of \he3 at 1 M$_\odot$ by a factor of 10.
For an initial deuterium abundance of 7.5 $\times 10^{-5}$, this
corresponds to a $g_3 = 0.27$. We reduced the degree to which the
Iben  \& Truran yields were modified at higher masses such that,
at M $>$ 3 M$_\odot$, we once again were using the Iben \& Truran
yields. The results of such a reduction in model 1 of Olive \etal (1995)
are shown in Figure 4. Here the evolution of D/H and \he3/H are shown
in model 1 with both the Iben \& Truran yields and the reduced yields.
Even in this case, there remains a mild overproduction of \he3 by a factor of
about 2.  That is, at $t = 9.4$ Gyr, \he3/H = 3 $\times 10^{-5}$.
In Figure 5, we show the effect of the reduced \he3 yields in the
model with outflow discussed in the previous section.
Here finally, we find a value for \he3/H at the solar epoch
which is perhaps acceptable. At $t = 9.4$ Gyr, \he3/H = 2.4 $\times 10^{-5}$.
A further improvement is possible by considering models which evolve on shorter
time scales as discussed above.

 An obvious problem with the reduction in \he3 yields at low stellar masses is
the observation of high $^3$He concentrations in planetary nebula
ejecta (Rood {\it \etal} 1992, 1995), which would seem to confirm the
predictions of the more standard models for the evolution of low
mass stars along the giant branch. It is clear that this issue must
be resolved before a more definitive statement can be made with respect
to Galactic evolution constraints on the primordial abundances of D and
$^3$He.

A further question of interest is that concerning the composition of
the matter involved in ``mass infall,'' during the later stages of evolution
of our Galaxy.
In this context, we note that while we have not considered such infall
models in this paper (see, e.g., Olive \etal 1995), they may provide
plausible alternative solutions to the \he3 problem.
The implications of infall of matter of primordial composition
of the light elements D. $^3$He, and $^7$Li are certainly quite different
from those of processed matter, which may generally be expected to be
metal enriched and deuterium depleted.
Infall of
primordial material is generally beneficial, with the adoption of the
lower primordial D abundance, while infall of D- and $^3$He-depleted matter
improves the situation for the case of a higher primordial D abundance.
The fact that the nature and origin of such infalling material is presently
uncertain, makes it necessary to treat its composition as an additional
parameter.
This problem is further
complicated by the fact that it is even possible
for the in falling gas both to be metal enriched and to
have an essentially primordial composition of D and $^3$He.
Such could occur if, for example, the ejecta of the first generation of
massive stars in the halo of our Galaxy were lost to the surrounding
intergalactic medium.
The ejecta of stars of M $>$ 10M$_\odot$ collectively represents
$\sim$ 10 \% of the initial mass formed into stars (e.g., for a Salpeter
IMF over the range 0.1-100 M$\odot$) and is characterized by a metal
abundance $>$ 10 Z$_\odot$. Assuming $\sim$ 10$^{10}$ M$_\odot$ of
early halo star formation would yield $\sim$ 10$^9$ M$_\odot$ of
metal enriched gas ejected, which could contaminate $\sim$ 10$^{10}$
M$_\odot$ to solar metallicity and yet have deuterium at a level of
only $\sim$ 0.9 its primordial value.

\section{Conclusions}

What can we conclude from this analysis?  We have shown that,
unlike the the abundances of some of the heavier elements such as
oxygen and neon which can differ by as much as
a factor of two locally relative to their average
galactic abundance by prior supernova in the solar neighborhood, the
local \he3 abundance could only have been affected by at most 10\%.
It also appears that the \he3 data from a variety of sources is consistent
and yields a value \he3/H$|_\odot = 1.5 \times 10^{-5}$ for the presolar
\he3 abundance.
Standard models of galactic chemical evolution yield an excess
of \he3 at the solar epoch by a factor which ranges from 2 to 12
depending on the assumed primordial value for D/H. For models
with D/H = $7.5 \times 10^{-5}$ initially, the factor nearly 4
excess in \he3 can be brought down to an excess of about 2
in models which favor massive stars early on, and include the possibility
for a substantial amount of metal enriched outflow.  In such models,
the solar \he3 abundance is brought down to nearly acceptable levels
when primordial D/H $< 3.5 \times 10^{-5}$.
Finally, we considered the possibility that part of the problem may
lie in the stellar yields of \he3. Though it appears that the cut in
yields suggested by Wasserburg \etal (1995) may not in itself be
sufficient to lower the solar \he3 abundance, that reduction in conjunction
with chemical evolution models may.
We feel justified in claiming that
any apparent ``crises" in big bang nucleosynthesis is rather
a (potential) problem for chemical evolution and/or stellar
evolution.

\bigskip
\noindent {\bf Acknowledgments}

We would like to thank  J. Audouze, C. Copi, R. Lewis, R. Pepin, G. Steigman
F. Timmes, and M. Turner for helpful
conversations.  The work of KAO was supported in part by DOE grant
DE-FG02-94ER-40823.  The work of DNS was supported in part by the DOE
(at Chicago and Fermilab) and by the NASA through NAGW-2381 (at
Fermilab) and a GSRP fellowship at Chicago.
The research of JWT has been supported by the
National Science Foundation under grant
NSF AST 92-17969 and by the National Aeronautics and Space Administration under
grant NASA NAG 5-2081.
The work of EV-F was
supported in part by PICS $n^o$114, ``Origin and Evolution of the
Light Elements", CNRS.

\newpage

\beginapjbib
\bibitem Bahcall, J.N. \& Pinsonneault, M.H. 1992, Rev. Mod. Phys.,
64,885

\bibitem Balser, D.S., Bania, T.M., Brockway, C.J.,
Rood, R.T., \& Wilson, T.L. 1994, ApJ, 430, 667

\bibitem Black, D.C. 1971, Nature Phys. Sci., 234, 148

\bibitem Black, D.C. 1972, Geochim. Cosmochim. Acta, 36, 347

\bibitem Bochsler, P. \& Geiss, J. 1989, in {\it Proc. Yosemite
Conf. on Outstanding Problems in the Solar System} p.133

\bibitem Bodmer, R., Bochsler, P., Geiss, J., Von Steiger, R.,
\& Gloeckler, G. 1995, Spa. Sci Rev., 72, 61

\bibitem Carswell, R.F., Rauch, M., Weymann, R.J., Cooke, A.J. \&
Webb, J.K. 1994, MNRAS, 268, L1

\bibitem Cass\'{e}, M., Olive, K.A., Scully, S.T., \& Vangioni-Flam, E.
1995, in preparation


\bibitem Charbonnel,C. 1994, A \& A, 282, 811

\bibitem Copi, C. Schramm,D.N. \& Turner, M.S. 1995a, Science,
267,192

\bibitem Copi, C. Schramm,D.N. \& Turner, M.S. 1995b, {ApJ Lett}
submitted

\bibitem Coplan, M.A., Ogilvie, K.W., Bochsler, P., \& Geiss, J.
1984, Solar Phys., 93, 415

\bibitem Cowan, J.J., Thielemann, F.-K., \& Truran, J.W. 1987, ApJ, 323, 543

\bibitem  Dearborn, D. S. P., Schramm, D.,
\& Steigman, G. 1986, ApJ, 302, 35

\bibitem Delbourgo-Salvador, P., Gry, C.,
Malinie, G., \& Audouze, J. 1985,
A\&A, 150, 53

\bibitem De Young, D.S. \& Heckman, T.M. 1994, ApJ, 431, 598

\bibitem Eberhardt, P. 1974, Earth Planet. Sci., 24, 182

\bibitem Edvardsson, B., Anderson, J., Gustafson, B.,
Lambert, D.L., Nissen, P.E., \& Tomkin, J. 1993, A \& A, 275, 101

\bibitem Elbaz, D., Arnaud, M., \& Vangioni-Flam, E. 1995, A \& A, in press


\bibitem Fields, B. 1995, ApJ, in press


\bibitem Galli, D., Palla, F. Ferrini, F., \& Penco,U. 1995, ApJ,
433, 536

\bibitem Geiss, J. 1993, in {\it Origin
 and Evolution of the Elements} eds. N. Prantzos,
E. Vangioni-Flam, and M. Cass\'{e}
(Cambridge:Cambridge University Press), p. 89

\bibitem Geiss, J. \& Reeves, H. 1972, A \& A, 18,126

\bibitem Grenon, M. 1989, Astrophys. \& Science, 156, 29

\bibitem Grenon, M. 1990, in {\it Astrophysical Ages and Dating Methods},
ed. E. Vangioni-Flam \etal (Ed. Frontieres, Paris), p. 153

\bibitem Gry, C., Malinie, G., Audouze, J.,
\& Vidal-Madjar, A. 1984, in
Formation and Evolution of Galaxies and Large
Scale Structure in the Universe,
eds. J. Audouze \& J. Tran Tranh Van (Reidel, Dordrecht) p 279



\bibitem Hata, N., Scherrer, R.J., Steigman, G., Thomas, D.,
Walker, T.P., Bludman, S., \& Langacker, P. 1995, preprint
hep-ph/9505319

\bibitem Hobbs, L. \& Thorburn, J. 1994, ApJ, 428, L25

\bibitem Hogan, C.J. 1995, ApJ, 441, L17

\bibitem Iben, I. 1967a, ApJ, 147, 624

\bibitem Iben, I. 1967b, ApJ, 147, 650

\bibitem Iben, I. \& Truran, J.W. 1978, ApJ, 220,980


\bibitem Larson, R.B. 1986, MNRAS, 218, 409

\bibitem Lattimer, J., Schramm, D.N., \& Grossman, L. 1977, ApJ, 214, 819

\bibitem Lee, T. 1979, Rev. Geophys. Space Phys., 17, 1591

\bibitem Levshakov, S.A.,  \& Takahara, F. 1995, preprint

\bibitem Linsky, J.L., Brown, A., Gayley, K., Diplas, A., Savage, B. D.,
Ayres, T. R., Landsman, W., Shore, S. N., Heap, S. R. 1993, ApJ, 402, 694

\bibitem Linsky, J.L.,  Diplas, A., Wood, B.E.,  Brown, A.,
Ayres, T. R.,  Savage, B. D., 1995, ApJ, in press

\bibitem Maeder, A. 1992, A \& A  264, 105

\bibitem Maeder, A. \& Meynet, G. 1994, A\&A, 287, 803

\bibitem Maeder, A., Lequeux, J., \& Azzopardi, M. 1980, A\&A, 90, L17

\bibitem Massey, P., Lang, C.C., DeGiola-Eastwood, K., \& Garmany, C.D. 1995,
ApJ, 438, 188

\bibitem McCray, R. \& Snow, T.P. 1979, ARAA, 17, 213

\bibitem Molaro, P., Primas, F., \& Bonifacio, P. 1995, A \& A, 295, L47

\bibitem Murphy, E.M., Lockman, F.J. \& Savage, B.D. 1995, ApJ, 447, 642

\bibitem Nissen, P.E. 1995, in IAU Symposium 164, {\it
Stellar Populations}, to be published

\bibitem Olive, K.A., Rood, R.T., Schramm, D.N., Truran, J.W.,
\& Vangioni-Flam, E. 1995, ApJ, 444, 680

\bibitem Olive, K.A., \& Schramm, D.N. 1981, ApJ, 257, 276

\bibitem Olive, K.A., \& Scully, S.T. 1995, Int. J. Mod. Phys. A,
in press

\bibitem Olive, K.A., \& Steigman, G. 1995, ApJ S, 97, 49

\bibitem Olsen, E.H. 1994, A\&A Supp, 104, 429


\bibitem Ostriker, J.P., \& Thuan, T.X. 1975, ApJ, 202, 353

\bibitem Ostriker, J.P., \& Tinsley, B. 1975, ApJ, 201, L51

\bibitem Pagel, B E.J., Simonson, E.A., Terlevich, R.J.
\& Edmunds, M. 1992, MNRAS, 255, 325

\bibitem Reeves, H. 1978, in {\it Protostars and Planets},
ed. T. Gehrels
(Tucson:University of Arizona Press)





\bibitem Rood, R.T., Bania, T.M., \& Wilson, T.L. 1992, Nature, 355, 618

\bibitem Rood, R.T., Bania, T.M.,  Wilson, T.L., \& Bania, D.S. 1995,
in {\it
 the Light Element Abundances, Proceedings of the ESO/EIPC Workshop},
ed. P. Crane, (Berlin:Springer), p. 201

\bibitem Rood, R.T., Steigman, G. \& Tinsley, B.M. 1976, ApJ, 207, L57

\bibitem Salpeter, E.E. 1955, ApJ, 121, 161

\bibitem Sasselov, D. \& Goldwirth, D.S. 1995, ApJ, 444, L5

\bibitem Scalo, J. 1986, Fund. Cosm. Phys. 11, 1


\bibitem Searle, L. \& Sargent, W.L. 1972, ApJ, 173, 25

\bibitem Skillman, E., \etal\ 1995, ApJ Lett (in preparation)

\bibitem Smith, V.V., Lambert, D.L., \& Nissen, P.E., 1992, ApJ 408, 262

\bibitem Smith, Scherpp \& Barnes 1989, ApJ, 336, 167

\bibitem Songaila, A., Cowie, L.L., Hogan, C. \& Rugers, M. 1994
Nature, 368, 599



\bibitem Steigman, G., Fields, B. D., Olive, K. A., Schramm, D. N.,
\& Walker, T. P., 1993, ApJ 415, L35



\bibitem Timmes, F.X., \& Truran, J.W. 1995, preprint

\bibitem Timmes, F.X., Woosley, S.E., \& Weaver, T.A. 1995, ApJSupp,
98, 617

\bibitem Tinsley, B.M. 1980, Fund. Cosmic Phys., 5, 287

\bibitem Tosi, M. 1988, A\&A, 197, 33

\bibitem Tosi, M., Steigman, G. \& Dearborn, D.S.P. 1995,  in {\it
 the Light Element Abundances, Proceedings of the ESO/EIPC Workshop},
ed. P. Crane, (Berlin:Springer), p.228

\bibitem Truran, J.W., \& Cameron, A.G.W. 1971, ApSpSci, 14, 179

\bibitem Turck-Chi\`eze, S., Cahen, S., Cass\'e, M., \&
Doom, C. 1988, ApJ, 335, 415

\bibitem Tytler, D. \& Fan, X.-M. 1995, BAAS,  26, 1424

\bibitem Vangioni-Flam, E., \& Audouze, J. 1988,
A\&A, 193, 81

\bibitem Vangioni-Flam, E. \& Cass\'{e}, M. 1995, ApJ, 441, 471

\bibitem Vangioni-Flam, E., Olive, K.A., \& Prantzos, N. 1994,
ApJ, 427, 618

\bibitem Vassiladis, E. \& Wood, P.R. 1993, ApJ, 413, 641

\bibitem Vidal-Madjar, A. 1991, Adv. Space Res., 11, 97

\bibitem Walker, T. P., Steigman, G., Schramm, D. N., Olive, K. A.,
\& Kang, H. 1991 ApJ, 376, 51

\bibitem Wang, Q.D., Walterbos, R.A.M., Steakley, M.F., Norman, C.A.,
\& Braun, R. 1995, ApJ, 439, 176

\bibitem Wasserburg, G.J., Boothroyd, A.I., \& Sackmann, I.-J. 1995, ApJ,
447, L37

\bibitem Weiler, R., Anders, E., Bauer, H., Lewis, R.
\& Signer, P. 1991, Geochim \& Cosmochim Acta, 55, 1709

\bibitem Weiss, A., Wagenhuber, J., and Denissenkov, P. 1995, preprint

\bibitem Woosley, S.E. \& Weaver, T.A. 1993, in {\it Supernova, Les Houches
Summer School Proceedings, Vol. 54}, ed. S. Bludman, R. Mochkovitch, \&
J. Zinn-Justin (Geneva: Elsevier Science Publishers), p. 100

\bibitem Wyse, R., \& Silk, J. 1987, ApJ, 313, L11


\endapjbib

\newpage
\noindent{\bf{Figure Captions}}

\vskip.3truein

\begin{itemize}

\item[]
\begin{enumerate}
\item[]
\begin{enumerate}

\item[{\bf Figure 1:}] The dependence of the metallicity produced
 as a function of the deuterium destruction
factor D/D$_p$, for a large sample of models. The metallicity is
plotted in solar units.  The SFR used was $\psi \propto M_{\rm gas}$
where the constant of proportionality ranges from 0.01 to 1.0
The circles correspond to the choice of stellar yields from Maeder (1992)
while the crosses correspond to the yields of Woosley \& Weaver (1993).
A power law IMF was chosen with a slope of -2.7 for
 the upper two sets of points and -3.0 for the lower two sets.

\item[{\bf Figure 2:}] The present day mass function of our adopted
model as compared with the data from Scalo (1986).

\item[{\bf Figure 3:}]  The evolution of
D/H and \he3/H
as a function of time, for a standard model
of galactic chemical evolution (solid line) and for one which
favors massive stars early and includes metal enriched outflow (dashed line).
Also shown are the values of these ratios at the time of formation of the sun,
$t \approx 9.4$ Gyr, and today, for
D/H (open squares) and \he3/H (filled circles).
The present day \he3 abundance simply shows the range of observed
values; the data point does not represent an average.
The  models were chosen
so that D/H is destroyed by a total factor of 5, to the present.

\item[{\bf Figure 4:}]  As in Figure 3, for a standard model
(solid) and for one in which
the stellar yields of \he3 at low masses have been reduced (dotted).
Deuterium is the same in both cases.

\item[{\bf Figure 5:}]  As in Figure 3, for the  model with enriched outflow
from Figure 3
(dashed) and for one with outflow in which
the stellar yields of \he3 at low masses have been reduced (dotted).

\end{enumerate}
\end{enumerate}
\end{itemize}

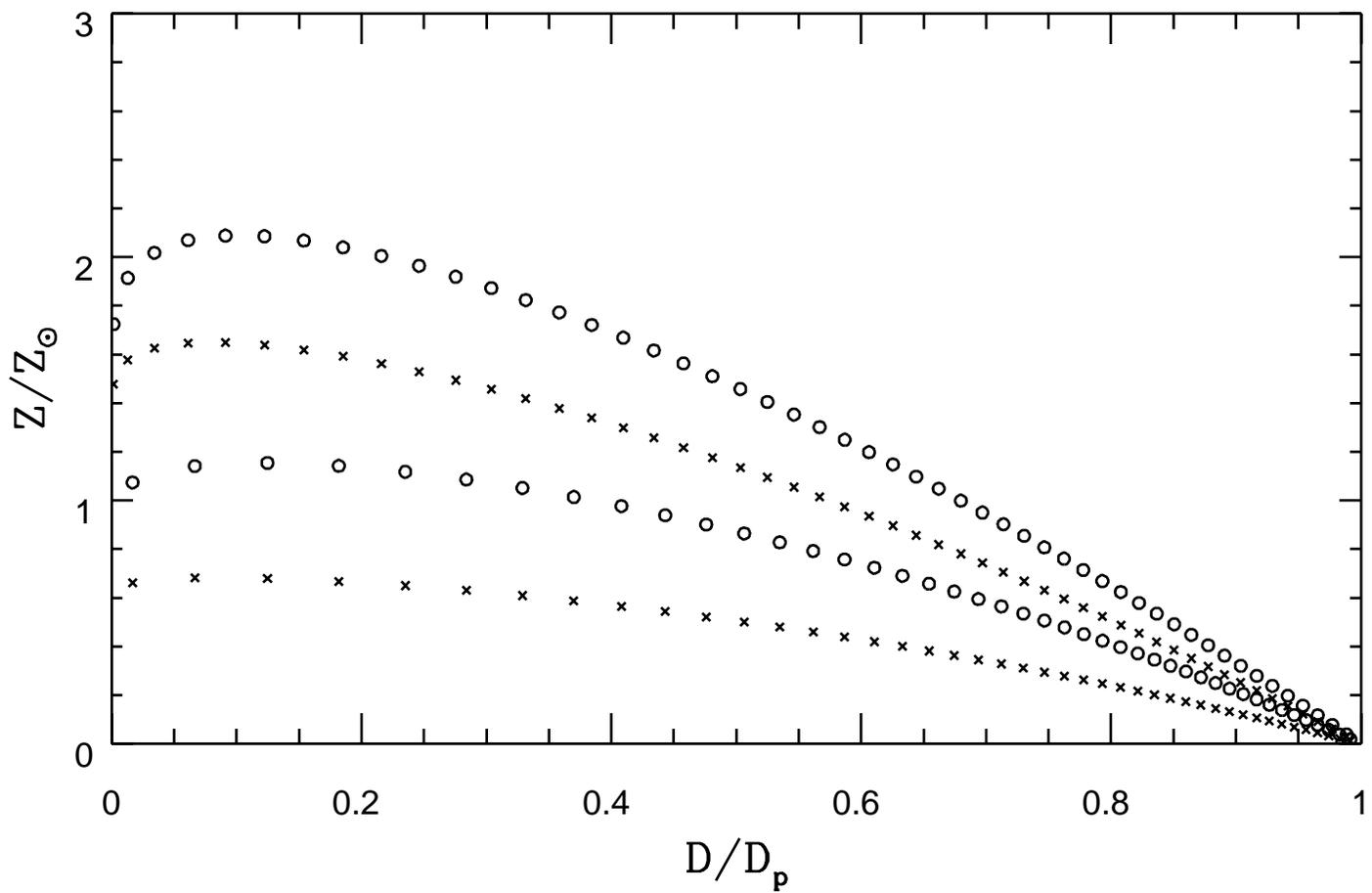
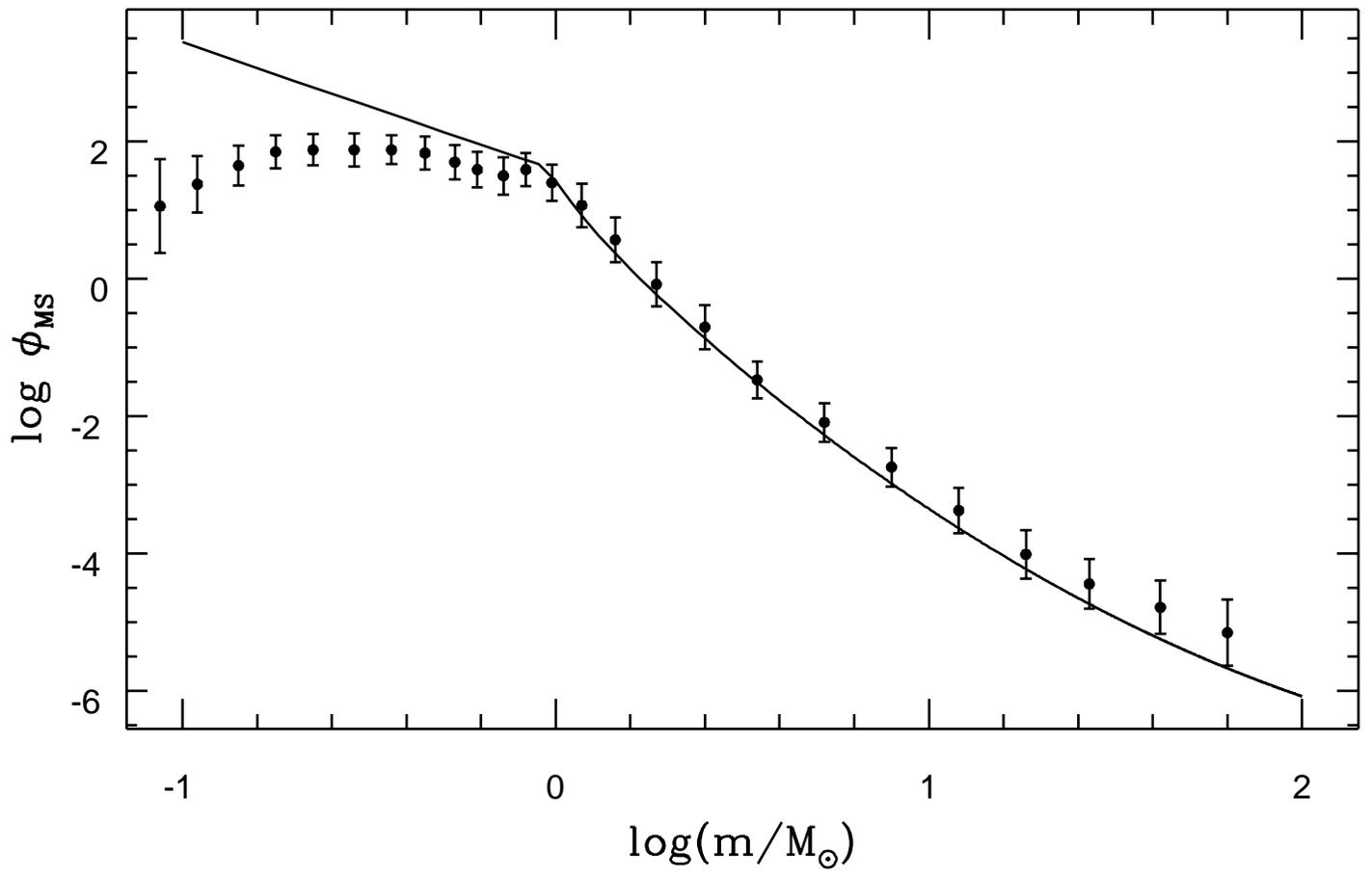
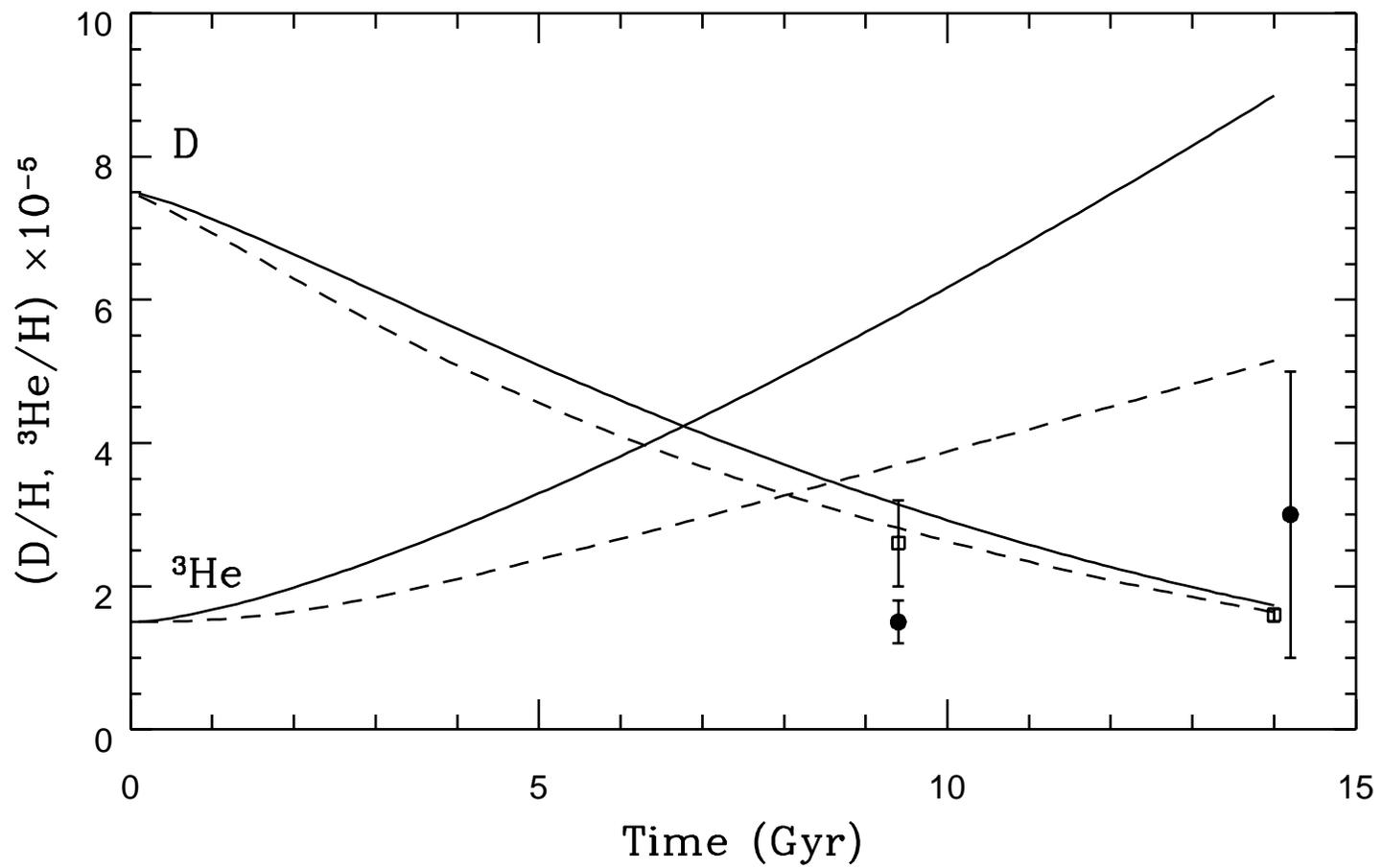
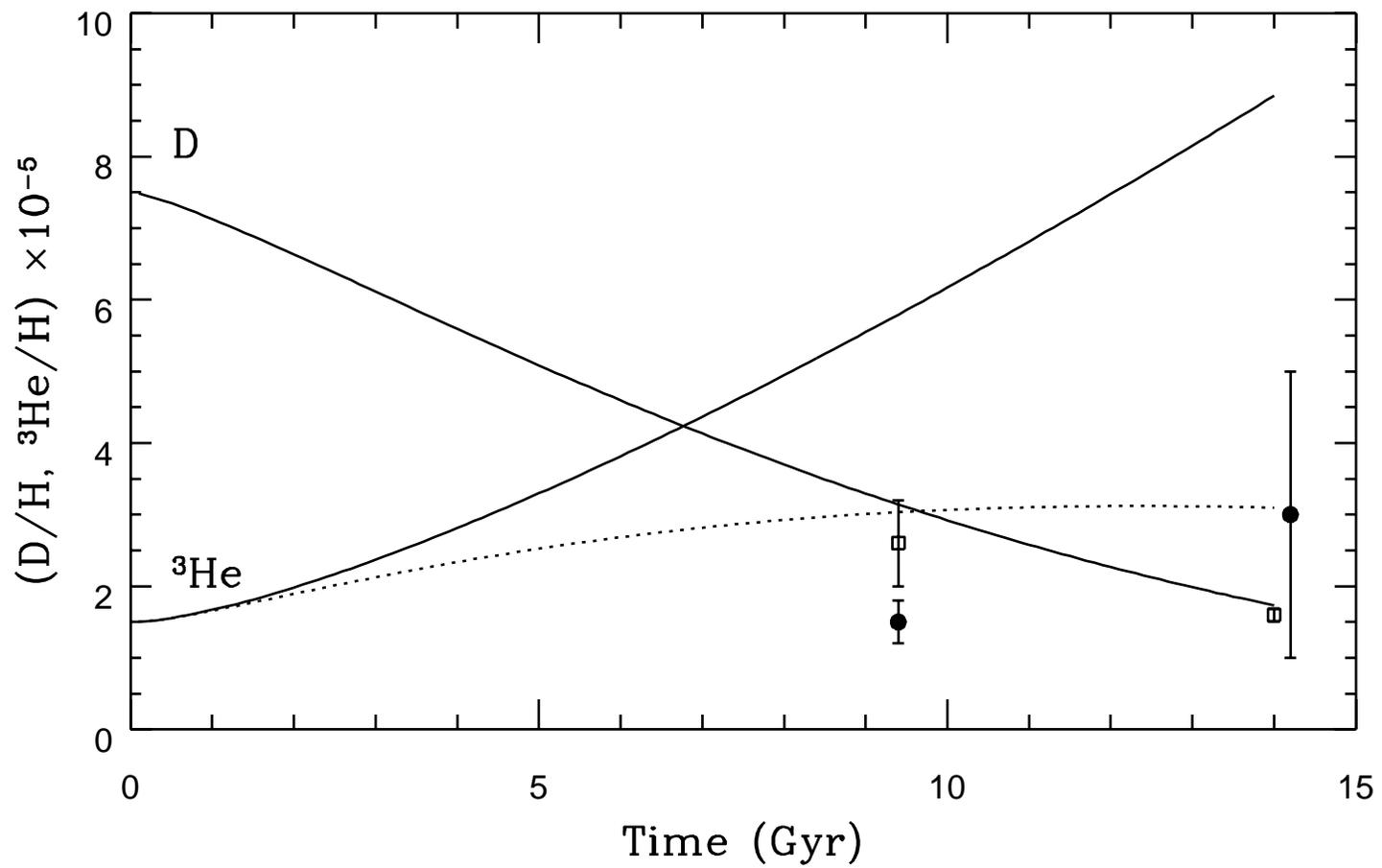
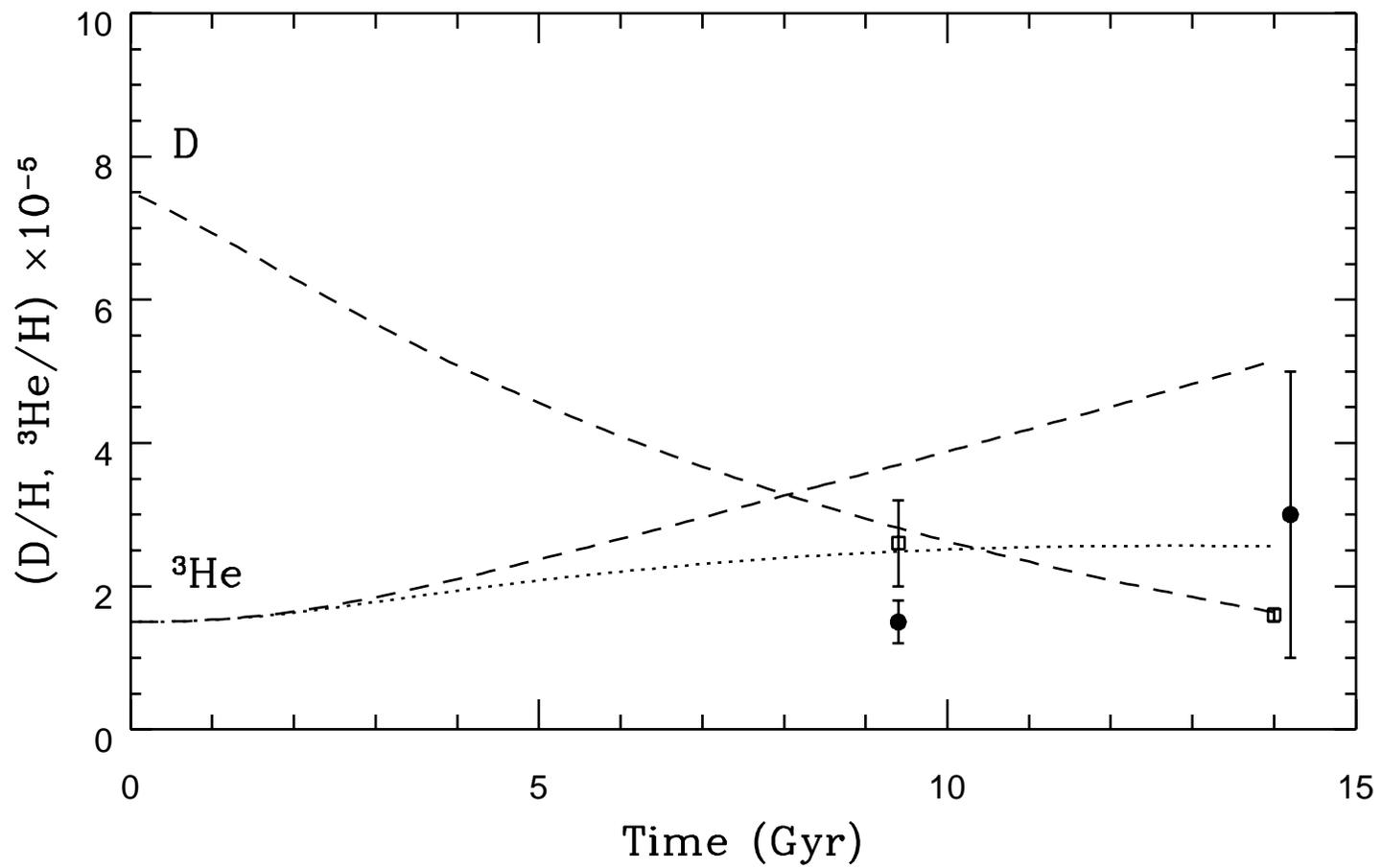
\end{document}